\documentclass[CEJP,PDF]{cej} 
\usepackage{layout}
\usepackage[OT4]{fontenc}
\usepackage{amsmath}
\usepackage{textcomp}
\usepackage{hyperref}
\usepackage{lscape}
\title{Isoscalar giant monopole resonance for drip-line
and super heavy nuclei in the framework
of a relativistic mean field formalism with scaling calculation
}

\articletype{Editorial}

\author{S. K. Biswal\email{sbiswal@iopb.res.in}\inst{ } 
        S. K. Patra\inst{ }
}

\institute{
     \inst{ } Institute Of Physics,\\
     Sachivalya Marg, 751 005, India.
          }
\date{\today}

\abstract{

We study the isoscalar giant monopole resonance for drip-lines and super heavy nuclei
in the frame work of a relativistic mean field theory with scaling approach. The
well known extended Thomas-Fermi approximation in the non-linear $\sigma$-$\omega$ model
is used to estimate the giant monopole excitation energy for some selected light spherical
nuclei starting from the region of proton to neutron drip-lines. The application is
extended to super heavy region for Z=114 and 120, which are predicted by several models
as the next proton magic number beyond Z=82.  We compared the excitation energy obtained
by four successful force parameters NL1, NL3, NL3$^*$ and FSUGold.
The monopole energy decreases toward the proton and neutron drip-lines in an isotopic
chain for lighter mass nuclei contrary to a monotonous decrease for super heavy isotopes.
The maximum and minimum monopole excitation energies are obtained for nuclei with
minimum and maximum isospin, respectively in an isotopic chain.
}

\begin{document}
\maketitle

\section{Introduction}

The study of nuclei far away from the drip-lines has a current research
interest due to their very different properties than the nuclei at the
$\beta-$stability valley. New properties of these nuclei like the soft giant
resonance, the change of magic number, the halo and skin structures and the
new decay modes stimulate strongly the research using radioactive ion beam (RIB)
\cite{tan85,han89}.
On the other hand, the super heavy nuclei which are on the stability line, but
extremely unstable due to the excessive Coulomb repulsion attract much theoretical
attention for its resemblance to the highly asymmetry nuclear matter limit \cite{og,kumar89}.
These nuclei possess a large amount of collective excitation and their study
along an isotopic chain is more informative for the structural evaluation of
astrophysical objects like neutron star \cite{aru04}.  Also, the nuclear symmetry
energy, and consequently the proton to neutron ratio, are
crucial factors in constructing the equation of state (EOS) for asymmetry nuclear
matter.

The compressibility $K_A$ of a nuclear system depends on its neutron-proton asymmetry. Also
it is well known that the EOS of an asymmetry accerating object like neutron star substantially
influence by it's compressibility. Although, the compressibility at various asymmetry is
an important quantities, it is not a direct experimental observable. Thus, one has to determine
the $K_A$ from the linked experimental quantity (which is directly or indirectly related
to $K_A$) like isoscalar giant monopole resonance (ISGMR)
\cite{young99,bohigas79}.
The ISGMR is a well defined experimental observable, which can be measured precisely through
various experimental techniques.  The drip-lines and super heavy nuclei are vulnerable
and unstable in nature because of the presence of excess neutron and large number of
protons, respectively. Thus it is instructive to know the giant monopole resonance,
compressibility modulus and other related quantities for both drip-lines and super heavy
nuclei. In this context, our aim is to study the giant monopole excitation energy
and the compressional modulus of finite nuclei near the drip-line \cite{han89} as well as
for recently discussed super heavy nuclei with proton numbers Z=114 and 120, which are
predicted to be next magic numbers beyond Z=82 with various models \cite{sil04,rutz95}.
In addition, the calculations of Refs. \cite{sobi,patra} suggest that these nuclei possess
spherical ground state or low-lying spherical excited solutions.
More specifically, we aimed to study the following within the frame-work of an
extended relativistic Thomas-Fermi approximation:

\begin{itemize}
\item
How the isoscalar excitation energy and the finite nuclear compressibility varies in an isotopic
chain in drip-lines and super heavy nuclei within a well tested model like relativistic
extended Thomas-Fermi frame work using scaling and constrained approaches which is developed
by some of us recently \cite{patra01,patra02}.
\end{itemize}
\begin{itemize}
\item
A comparative study of ISGMR obtained with various parameter sets  such as
NL1, NL3, NL3$^*$ and FSUGold for the same drip-lines and super heavy nuclei.
\end{itemize}
\begin{itemize}
\item
The resonance width $\Sigma$, which is mostly the difference between the scaling and constraint
excitation energies are analyzed in the isotopic chains of light and super heavy nuclei.

\end{itemize}
\begin{itemize}
\item
Finally, the relation between the finite nuclear compressibility with the infinite nuclear matter values in various force parameter sets are looked for.
\end{itemize}

In relativistic mean field (RMF) formalism the NL1 parameter set
\cite{rufa86} is considered to be
one of the best interaction for a long time to predict the experimental observables. The
excessive large value of asymmetry coefficient $J\sim 43.6$ MeV questions about the
accuracy for the prediction
of neutron radius near the drip-line. As a result, the discovery of NL3 parameter set
\cite{lala97} complement the limitation of NL1 force and evaluates the
ground state properties of finite nuclei in an excellent agreement
with experiment \cite{boguta87,lala97,patra91,gamb90,sumi93,toki94}. It reproduces
the proton  or charge radius $r_c$ precisely along with the ground state binding
energy. Unfortunately, the experimental data for neutron
radius has a large error bar \cite{batty}, which covers most of the prediction of all
relativistic and non-relativistic models \cite{brow00}. The
FSUGold parameter set \cite{todd05,fattoyev10} reproduces the ISGMR pretty well
with the experimental data for $^{90}$Zr and $^{208}$Pb. There is also a possibility to
solve the uncertainty in neutron radius problem \cite{roca11} using this interaction.
The NL3$^*$ force parametrization \cite{lala09} claimed to be an improved version
of NL3 to reproduce the
experimental observables. We used all these forces and made a comparison
of their predictive power for various experimental data. Then we selected NL3 as a suitable
parameter set for our further investigations for ISGMR and related quantities.

The paper is presented as follows: In section II, we outline
in brief the formalism used in the present work.  In section III,
we discuss our results for the ground state isoscalar giant monopole
resonance (ISGMR)
for drip-lines and super heavy nuclei.
The isoscalar monopole excitation energy $E_x$ and the compressibility modulus
of finite nuclei $K_A$ are also analyzed. We give the summary and
concluding remarks in section IV.

\section{The Formalism}

In this paper we shall make use of the principle of scale invariance
to obtain the virial theorem for the relativistic mean field
\cite{serot86} theory by working in the
relativistic Thomas--Fermi (RTF) and relativistic extended Thomas-Fermi
(RETF) approximations \cite{patra01,mario93,mario92,spei98,mario98,mario93a}.
Although, the scaling and constrained calculations are not new, the present
technique is developed first time by Patra et al \cite{patra01,mario10} and
not much have been explored for various regions of the periodic chart.
Thus, it is interesting to apply the model specially for drip-lines and
super heavy nuclei.
The calculations will be explored to the region
ranging from Z=8 to Z=114, 120, where we can simulate the
properties of neutron matter from the neutron-rich finite nuclei.
For this purpose we compute moments
and average centroid energies of the isoscalar giant monopole resonance
(ISGMR) through scaling and constrained self-consistent calculations for
ground state.

The detail formalisms of the scaling method are given in Refs. \cite{patra01,patra02}.
For completeness, we have outlined briefly some of the essential expressions
which are needed for the present purpose. We have worked  with the non-linear
Lagrangian of Boguta and Bodmer \cite{boguta77} to include the many-body correlation
arises from the non-linear terms of the $\sigma-$meson self-interaction \cite{schiff51}
for nuclear many-body system. The nuclear matter compressibility modulus $K_{\infty}$ also
reduces dramatically by the introduction of these terms, which
motivates to work with this non-liner Lagrangian.
The relativistic mean field Hamiltonian for a nucleon-meson interacting system
is written by \cite{serot86,patra01}:
\begin{eqnarray}
{\cal H}&= &\sum_i \varphi_i^{\dagger}
\bigg[ - i \vec{\alpha} \cdot \vec{\nabla} +
\beta m^* + g_{v} V + \frac{1}{2} g_{\rho} R \tau_3 \nonumber\\
&+&\frac{1}{2} e {\cal A} (1+\tau_3) \bigg] \varphi_i 
+ \frac{1}{2} \left[ (\vec{\nabla}\phi)^2 + m_{s}^2 \phi^2 \right]
+\frac{1}{3} b \phi^3\nonumber\\
&+& \frac{1}{4} c \phi^4
-\frac{1}{2} \left[ (\vec{\nabla} V)^2 + m_{v}^2 V^2 \right]\nonumber\\
&-& \frac{1}{2} \left[ (\vec{\nabla} R)^2 + m_\rho^2 R^2 \right]
- \frac{1}{2} \left(\vec{\nabla}  {\cal A}\right)^2 . 
\end{eqnarray}
Here $m$, $m_s$, $m_v$ and $m_{\rho}$ are the masses for the
nucleon (with $m^*=m-g_s\phi$ being the effective mass of the nucleon),
${\sigma}-$, $\omega-$ and ${\rho}-$mesons, respectively
 and ${\varphi}$ is the Dirac spinor.
The field for the ${\sigma}$-meson is denoted by ${\phi}$, for ${\omega}$-meson
by $V$, for ${\rho}$-meson by $R$ ($\tau_3$ as the $3^{rd}$ component of the
isospin)  and for photon by $A$.
$g_s$, $g_v$, $g_{\rho}$ and $e^2/4{\pi}$=1/137 are the coupling
constants for the ${\sigma}$, ${\omega}$, ${\rho}$-mesons and photon respectively.
$b$ and $c$ are the non-linear coupling constants for ${\sigma}$ mesons.
By using the classical variational principle we obtain the field equations for
the nucleon and mesons. In semi-classical approximation we can write the above
Hamiltonian in term of density as:
\begin{eqnarray}
{\cal H}&=&{\cal E}+g_v V {\rho}+g_{\rho}R{\rho}_3+e{\cal A}{\rho}_p+{\cal H}_f,
\end{eqnarray}
where
\begin{eqnarray}
{\cal E}&= &\sum_i \varphi_i^{\dagger}
\bigg[ - i \vec{\alpha} \cdot \vec{\nabla} +
\beta m^*\bigg]\varphi_i,
\end{eqnarray}
\begin{eqnarray}
{\rho}_s&=&\sum_i \varphi_i^{\dagger}{\varphi},
\end{eqnarray}
\begin{eqnarray}
{\rho}&=&\sum_i {\bar \varphi_i}{\varphi},
\end{eqnarray}
\begin{eqnarray}
{\rho_3}&=& \frac {1}{2}\sum_i{\varphi}_i^{\dagger}{\tau_3}{\varphi}_i,
\end{eqnarray}
and ${\cal H}_f$ is the free part of the Hamiltonian.
The total density $\rho$ is the sum of proton $\rho_p$ and neutron
$\rho_n$ densities.
The semi-classical ground-state meson fields are obtained
by solving the Euler--Lagrange equations $\delta {\cal H}/\delta \rho_q = \mu_q$
($q=n, p$).
\begin{equation}
(\Delta- m_s^2)\phi = -g_{s} \rho_{s} +b\phi^2 +c\phi^3 ,
\label{eqFN4}  \\[3mm]
\end{equation}
\begin{equation}
(\Delta - m_{v}^2) V =  -g_{v} \rho,
\label{eq18} 
\end{equation}
\begin{equation}
(\Delta - m_\rho^2)  R  =   - g_\rho \rho_3,
\label{eq19} 
\end{equation}
\begin{equation}
\Delta {\cal A}    =      -e \rho_{p}.
\label{eqFN7}
\end{equation}
The above field equations are solved self-consistently in an iterative method.
The pairing correlation is not included in the evaluation of the equilibrium property
as well as monopole excitation energy. The Thomas-Fermi approach is a semi-classical
approximation and pairing correlation has a minor role in giant resonance.
It is shown in \cite{xavier11,baldo13} that only for open-shell nuclei,
it has a marginal effect on ISGMR energy.  As far as pairing correlation is concerned,
it is a quantal effect and  can be included in a semi classical calculation as an
average as it is adopted in semi-empirical mass formula.  In Ref.
\cite{baldo13},  perturbative
calculation on top of a semi classical approach is done, and it suggests that pairing
correlation is unimportant in such approach like relativistic Thomas-Fermi
(RTF) or relativistic extended Thomas-Fermi (REFT) approximations.  In our present
calculations, the scalar density ($\rho_s$) and energy density ($\cal{E}$) are calculated
using RTF and RETF formalisms. The RETF is the ${\hbar}^2$ correction to the RTF, where the
gradient of density is taken care. This term of the density takes care of the
variation of the density and involves more in the surface properties.
\begin{multline}
{\cal H} = {\cal E} + \frac{1}{2}g_{s}\phi \rho^{eff}_{s}
+ \frac{1}{3}b\phi^3+\frac{1}{4} c \phi^4
+\frac{1}{2} g_{v}  V \rho +\frac{1}{2} g_\rho R \rho_3 
+\frac{1}{2} e {\cal A} \rho_{p},
\label{eqFN8c}
\end{multline}
with
\begin {eqnarray}
{\rho}_s^{eff}&= & g_s{\rho}_s-b{\phi}^2-c{\phi}^3.
\end{eqnarray}
In order to study the monopole vibration of the nucleus we have scaled the baryon density
\cite{patra01}.
The normalized form of the baryon density is given by
\begin {eqnarray}
{\rho}_{\lambda}\left(\bf r \right)&=&{\lambda}^3{\rho}\left(\lambda r \right),
\end{eqnarray}
${\lambda}$ is the collective co-ordinate associated
 with the monopole vibration. As Fermi momentum and
 density are inter-related, the scaled Fermi momentum is given by
\begin {eqnarray}
K_Fq{\lambda}&=&\left[3{\pi}^2{\rho}_q
\lambda\left(\bf r \right)\right]^{\frac{1}{3}}.
\end {eqnarray}
Similarly $\phi$, $V$, $R$ and Coulomb fields are scaled due to self-consistence
eqs. (7-10).
But the $\phi$ field can not be scaled simply like the density and momentum,
 because the source term of $\phi $  field contains the $\phi$ field itself. In
semi-classical formalism, the energy and density are scaled like
\begin{eqnarray}
{\cal E_{\lambda}}(\bf r)&=&{\lambda}^4 {\tilde{\cal E}}(\lambda \bf r)\nonumber\\
&=& \lambda^4[\tilde{\cal E}_{0}(\lambda \bf r)+
\tilde{\cal E}_2(\lambda \bf r)],
\end {eqnarray}
\begin{eqnarray}
\rho_{{s\lambda}}(\bf r)={\lambda^3}{\tilde{\rho}}_{s}{(\lambda{\bf r})}.
\end{eqnarray}
The symbol $\sim$ shows an implicit dependence of $ \tilde m^* $.
With all these scaled variables, we can write the Hamiltonian as:
\begin{eqnarray}
{\cal{H}}_{\lambda} & = & {\lambda^3}{\lambda}{\tilde{\cal{E}}}+\frac{1}{2}{g_s}
\phi_{\lambda}{{\tilde{\rho}}_s^{eff}}\nonumber
+\frac{1}{3}\frac{b}{\lambda^3}\phi_\lambda^3
+\frac{1}{4}\frac{c}{\lambda^3}{\phi_\lambda}^4\\
&+&\frac{1}{2}{g_v}V_\lambda{\rho}
+\frac{1}{2}g_{\rho}R_\lambda {\rho_3}+\frac{1}{2}e{A}_\lambda{\rho}_p.
\end{eqnarray}
Here we are interested to calculate the monopole excitation energy which is defined as
${E}^{s}={\sqrt{\frac{C_m}{B_m}}}$ with ${C_m}$ is the restoring force and
$B_m$ is the mass parameter. In our calculations, $C_m$ is
obtained from the double derivative of the scaled energy with respect to the scaled
co-ordinate $\lambda$ at ${\lambda}=1$ and is defined as \cite{patra01}:
\begin {eqnarray}
{C}_m&=&\int{dr}\bigg[-m\frac{\partial{\tilde{\rho_s}}}{\partial{\lambda}}
+3\bigg({m_s}^2{\phi}^2+\frac{1}{3}b{\phi}^3
\nonumber\\
&-& {m_v}^2{V^2}
-{m_{\rho}}^2R^2\bigg)-(2{m_s}^2{\phi}
+b{\phi}^2)\frac{\partial{\phi_\lambda}}{\partial{\lambda}}\nonumber\\
&+&2{m_v}^2V\frac{\partial{V_\lambda}}{\partial{\lambda}}
+2{m_{\rho}}^2R
\frac{\partial R_\lambda}{\partial \lambda}\bigg]_{\lambda=1},
\end{eqnarray}
and the mass parameter $B_{m}$ of the monopole vibration can be expressed as the
double derivative of the scaled energy with the collective velocity $\dot{\lambda}$
as
\begin{eqnarray}
B_{m}=\int{dr}{U(\bf r)}^2{\cal {H}},
\end {eqnarray}
where $U(\bf r)$ is the displacement field, which can be determined from the relation
between collective velocity $\dot{\lambda}$ and velocity of the moving frame,
\begin {eqnarray}
U(\bf r)=\frac{1}{\rho(\bf r){\bf r}^2}\int{dr'}{\rho}_T(r'){r'}^2,
\end {eqnarray}
with ${\rho}_T $ is the transition density defined as
\begin {eqnarray}
{{\rho}_T(\bf r)}=\frac{\partial{\rho_\lambda(\bf r)}}{\partial{\lambda}}\bigg|_{\lambda = 1}
=3 {\rho}(\bf r)+r \frac{\partial{\rho(\bf r)}}{\partial r},
\end {eqnarray}
taking $U(\bf r)=r $. Then the mass parameter can be written as
$B_m=\int{dr}{r}^2{\cal H}$.
In non-relativistic limit, ${B_m}^{nr}=\int{dr}{r^2}m{\rho}$ and the
scaled energy ${E_m}^{s}$ is $\sqrt{\frac{m_3}{m_1}}$.
The expressions for ${m_3}$ and ${m_1}$
can be found in \cite{bohigas79}.
Along with the scaling calculation, the monopole vibration can
also be studied with constrained approach \cite{bohigas79,maru89,boer91,stoi94,stoi94a}.
In this method, one has to solve the
constrained functional equation:
\begin{eqnarray}
\int{dr}\left[{\cal H}-{\eta}{r}^2 {\rho}\right]=E(\eta)-\eta\int{dr}{r}^2\rho.
\end{eqnarray}
Here the constrained is ${\langle {R^2}\rangle}_0 ={\langle{r^2}\rangle}_m$. The
constrained energy $E(\eta)$ can be expanded in a harmonic approximation as
\begin{eqnarray}
E(\eta)& = &E(0)+\frac{\partial E(\eta)}{\partial \eta}\big|_{\eta =0}
+\frac{\partial^2{E(\eta)}}{\partial{\eta}^2}|_{\eta =0}.
\end {eqnarray}
The second order derivative in the expansion is related with the constrained
compressibility modulus for finite nucleus $K_A^c$ as
\begin{eqnarray}
{K_A}^{c} = \frac{1}{A} {R_0}^{2} \frac{\partial^2{E \eta}}
{\partial{R_\eta}},
\end{eqnarray}
and the constrained energy ${{E_m}^{c}}$ as
\begin {eqnarray}
{{E_m}^{c}}={\sqrt{\frac{A {K_A^c}}{B_m^c}}}.
\end {eqnarray}
In the non-relativistic approach, the constrained energy is related by the
sum rule weighted ${{E_m}^{c}}={\sqrt{\frac{m_1}{m_{-1}}}}$. Now the scaling and constrained
excitation energies of the monopole vibration in terms of the non-relativistic
sum rules will help us to calculate $\Sigma$, i.e., the resonance width \cite{bohigas79,mari05},
\begin{eqnarray}
{\Sigma} &=& \sqrt{\left({E_m}^s\right)^2-\left({E_m}^c\right)^2} 
= \sqrt{({\frac{m_3}{m_1}})^2-{(\frac{m_1}{m_{-1}}})^2}. 
\end{eqnarray}

\section{Results and Discussions}

\begin{table}
\caption{The calculated binding BE and charge radius $r_c$ obtained from
relativistic extended Thomas-Fermi (RETF) approximation is compared with relativistic
Hartree (with various parameter sets) and experimental results \cite{wang12,angeli13}.
The  RETF results
are given in the parenthesis. The empirical values \cite{bethe71,blaizot}
of nuclear matter saturation
density $\rho_0$,
binding energy per nucleon BE/A, compressibility modulus $K$,
asymmetry parameter $J$ and ratio of effective mass to the nucleon mass
$M^*/M$ are given in the
lower part of the table. The energy is in MeV and radius is in fm.
}
\renewcommand{\tabcolsep}{0.02cm}
\renewcommand{\arraystretch}{1.0}
\begin{tabular}{|ccccccc|}
\hline
\hline
Nucleus & Set     & BE (calc.)  & BE (Expt.) & $r_c$ (calc.) & $r_c$ (Expt.)& \\
\hline
$^{16}$O & NL1 & 127.2(118.7) &127.6  & 2.772(2.636) &2.699  & \\
         & NL3 & 128.7(120.8) &  & 2.718(2.591) &  & \\
         & NL3$^*$& 128.1(119.5) &  & 2.724(2.603) &  & \\
         & FSUG& 127.4(117.8) &  & 2.674(2.572) &  & \\
$^{40}$Ca& NL1 & 342.3(344.783) &342.0  & 3.501(3.371) & 3.478 & \\
         & NL3 & 341.6(346.2) &  & 3.470(3.343) &  & \\
         & NL3$^*$& 341.5(344.2) &  & 3.470(3.349) &  & \\
         & FSUG& 340.8(342.2) &  & 3.429(3.327) &  & \\
$^{48}$Ca& NL1 & 412.7(419.5) & 416.0 & 3.501(3.445) & 3.477 & \\
         & NL3 & 414.6(422.6) &  & 3.472(3.426) &  & \\
         & NL3$^*$& 413.5(420.3) &  & 3.469(3.429) &  & \\
         & FSUG& 411.2(418.0) &  & 3.456(3.418) &  & \\
$^{90}$Zr& NL1 & 784.3(801.1) &783.9 & 4.284(4.232) & 4.269 & \\
         & NL3 & 781.4(801.7) &  & 4.273(4.219) &  & \\
         & NL3$^*$& 781.6(798.7) &  & 4.267(4.219) &  & \\
         & FSUG& 778.8(797.3) &  & 4.257(4.214) &  & \\
$^{116}$Sn&NL1 & 989.5(1013.7) &988.7  & 4.625(4.583) & 4.625 & \\
          &NL3 & 985.4(1014.6) &  & 4.617(4.571) &  & \\
         & NL3$^*$& 986.4(1011.0) &  & 4.609(4.569) &  & \\
         & FSUG& 984.4(1010.7) &  & 4.611(4.569) &  & \\
$^{208}$Pb&NL1 & 1638.1(1653.7)&1636.4  & 5.536(5.564) & 5.501 & \\
          &NL3 & 1636.9(1661.2)&  & 5.522(5.541) &  & \\
         & NL3$^*$& 1636.5(1655.2)&  & 5.512(5.538) &  & \\
         & FSUG& 1636.2(1661.4)&  &5.532(5.541)  &  & \\
\hline
\hline
Set &NL1& NL3&NL3$^*$&FSUG &emperical& \\
\hline
\hline
$\rho_0$ &0.154 & 0.150 &0.148 &0.148&0.17 &  \\
$E/A$ &16.43&16.31&16.30&16.30&15.68&\\
$K$ &211.7&271.76&258.27&230.0&$210\pm 30$&\\
$J$ &43.6&38.68&37.4&32.597&$32\pm 2$&\\
$M^*/M$ &0.57& 0.594&0.60&0.61&0.6&\\
\hline
\end{tabular}
\label{table1}
\end{table}

\subsection{Force parameter of relativistic mean field formalism}

First of all, we examined the predictive power of various parameter sets. In this
context we selected NL1 as a successful set of past and few recently used forces
like NL3, NL3$^*$ and FSUGold. The ground state observables obtained by these forces
are depicted in Table \ref{table1}. Along with the relativistic extended
Thomas-Fermi (RETF) results, the
values with relativistic Hartree are also compared with the experimental data
\cite{wang12,angeli13}. The calculated RMF results obtained by all the force parameters
considered in the present paper are very close to the experimental data \cite{wang12,angeli13}.
A detail analysis of the binding energy and charge radius clear that NL1 and FSUGold
have a superior predictive power for $^{16}$O in RMF level. The advantage of
FSUGold decreases with increase of mass number of the nucleus. Although, the
predictive power of the pretty old NL1 set is very good for binding energy, it has a
large asymmetry coefficient $J$, which may mislead the prediction in unknown
territory, like neutron drip-line or super heavy regions. The RETF prediction of
binding energy and charge radius (numbers in the parenthesis) is very poor
with the experimental data as compared to the RMF calculations. However, for
relatively heavier masses, the ERTF results can be used within
acceptable error.  In general, taking into account the
binding energy BE and root mean square charge radius $r_c$, one may prefer to use
either NL3 or NL3$^*$ parametrization.
\begin{table*}
\caption{The results of isoscalar giant monopole resonance with
various parameter sets for some known nuclei are compared with recent experimental
data \cite{young13}. The calculations are done with
relativistic extended Thomas-Fermi (RETF) approximation using both
scaling and constrained schemes. The values of $\Delta_1$,
$\Delta_2$ and $\Delta_3$ are obtained by subtracting the
results of (NL3, NL3$^*$), (NL3$^*$, FSUG) and (NL3, FSUG), respectively.
The monopole excitation energies with scaling $E^s$ and constrained $E^c$
are in MeV.
}
\renewcommand{\tabcolsep}{0.30cm}
\renewcommand{\arraystretch}{1.0}
\begin{tabular}{|c|c|c|c|c|c|c|c|c|c|c|c|c|c|}
\hline
\hline
Nucleus
& \multicolumn{2}{c|}{NL1}
& \multicolumn{2}{c|}{NL3}
&\multicolumn{2}{c|}{NL3$^*$}
&\multicolumn{2}{c|}{FSUG}
&\multicolumn{1}{c|}{Expt.}
&{$\Delta_1$}&{$\Delta_2$}&{$\Delta_3$} \\
\hline
& ${E}^s$& ${E}^c$&${E}^s$& ${E}^c$& ${E}^s$& ${E}^c$& ${E}^s$& ${E}^c$& &
 && \\
\hline
$^{16}{O}$&23.31&21.75&27.83&25.97&26.86&25.20&26.97&25.17&21.13$\pm$0.49&0.97&0.11&0.86 \\
$^{40}{Ca}$&20.61&19.77&24.01&23.16&23.32&22.48&22.98&22.30&19.20$\pm$0.40&0.69&0.43&1.03\\
$^{48}{Ca}$&19.51&18.67&22.69&21.73&22.01&21.11&21.72&20.88&19.90$\pm$0.20&0.68&0.29&0.97\\
$^{90}{Zr}$&16.91&16.41&19.53&19.03&18.97&18.50&18.60&18.21&$17.89\pm 0.20$&
0.56&0.37&0.97\\
$^{110}{Sn}$&15.97&15.50&18.42&17.94&17.90&17.44&17.52&17.13&&0.52&0.38&0.90\\
$^{112}{Sn}$&15.87&15.39&18.29&17.81&17.78&17.32&17.42&17.02&16.1$\pm$0.10&0.51&0.36&0.86\\
$^{114}{Sn}$&15.76&15.28&18.16&17.67&17.65&17.18&17.31&16.90&15.9$\pm$0.10&0.51&0.34&0.85\\
$^{116}{Sn}$&15.63&15.19&18.02&17.52&17.51&17.04&17.19&16.77&15.80$\pm$0.10&0.51&0.32&0.83\\
$^{118}{Sn}$&15.51&15.03&17.87&17.36&17.37&16.89&17.07&16.63&15.6$\pm$0.10&0.50&0.30&0.80\\
$^{120}{Sn}$&15.38&14.90&17.72&17.20&17.22&16.73&16.94&16.49&15.4$\pm$0.20&0.50&0.28&0.78\\
$^{122}{Sn}$&15.24&14.76&17.56&17.03&17.07&16.57&16.81&16.34&15.0$\pm$0.20&0.49&0.24&0.77\\
$^{124}{Sn}$&15.11&14.61&17.40&16.85&16.91&16.40&16.67&16.19&14.80$\pm$0.20&0.48&0.24&0.72\\
$^{208}{Pb}$&12.69&12.11&14.58&13.91&14.18&13.55&14.04&13.44&14.17$\pm$0.28&0.40&0.14&0.54\\
$^{286}{114}$&11.32&10.60&13.00&12.14&12.64&11.83&12.55&11.79&&0.36&0.09&0.45\\
$^{298}{114}$&11.05&10.31&12.68&11.80&12.33&11.50&12.29&11.53&&0.35&0.04&0.37\\
$^{292}{120}$&11.28&10.53&12.96&12.07&12.60&11.76&12.48&11.69&&0.36&0.12&0.27\\
$^{304}{120}$&11.04&10.28&12.67&11.77&12.33&11.47&12.25&11.47&&0.34&0.08&0.42\\
\hline
\hline
\end{tabular}
\label{table2}
\end{table*}
Before accepting NL3 or NL3$^*$ as the usuable parameter set for our further calculations,
in Table \ref{table2}, we have given the excitation energy of some selective
nuclei both in light and super heavy regions with various parameter sets for some
further verification. The
isoscalar giant monopole energies $E^s$ and $E^c$ are evaluated using both scaling and
constraint calculations, respectively. The forces like NL1, NL3, Nl3$^*$ and  FSUG  have a
wide range of compressibility $K_{\infty}$ starting from 211.7 to 271.7 MeV
(see Table \ref{table1}).
Because of the large variation in $K_{\infty}$ of these sets, we expect various values of
$E^s$ and $E^c$ with different parametrization. From Table
\ref{table2}, it is noticed that the calculated results for $^{16}$O and $^{40}$Ca differ
substantially from the data. Again this deviation of calculated result goes on decreasing
with increase of mass number, irrespective of the parameter set. This may be
due to the  use of semi-classical approximation like Thomas-Fermi  and extended Thomas-Fermi.
In these approaches, quantal corrections are averaged out. When we are going from light to
the heavy and then super heavy nuclei, the surface correction decreases appreciably.
Consequently, the contribution to monopole excitation energy decreases with mass number A.
In column 11, 12 and 13 of Table \ref{table2}, the differences in $E_x$ obtained from
various parameter sets are given, namely, $\triangle_{1}$ is the difference in monopole
excitation energy obtained by NL3 and NL3$^*$. Similarly, $\triangle_{2}$
and $\triangle_{3}$ are the ISGMR difference with  (NL3$^*$, FSUG) and (NL3, FSUG), respectly.
The values of $\triangle_{1}$, $\triangle_{2}$ or  $\triangle_{3}$ goes on decreasing
with increase of mass number of the nucleus without depending on the
parameter used. In other word, we may reach to same conclusion in the super heavy region
irrespective of the parameter set.
However, it is always better to use a successful
parameter set to explore an unknown territory. In this context, it is safer to
choose NL3 force for our further exploration.
The second observation is also apparent from the Table. It is commonly believe that, mostly
the compressibility of the force parameter affect
the excitation energy of ISGMR of the nucleus. That means force parameters having different
$K_{\infty}$ have different excitation energy for the same nucleus. For example,
$^{208}{Pb}$ has excitation energy 14.58 and 14.04 MeV with NL3 and FSUGold, respectively.
Although, the ground state binding energy of $^{208}$Pb, either with
Hartree (RMF) or REFT approximation matches well with  NL3 and FSUGold parameter sets
(see Table \ref{table1}), their ISGMR differ by 0.54 MeV, which is quite substantial. This
unequal prediction of $E^s$ may be due to the difference in nuclear matter compressibility
of the force parametrizations.
\begin{table*}
\caption{The predicted proton and neutron drip-lines PDL and NDL for O, Ca, Ni, Sn, Pb,
Z=114 and Z=120 in relativistic mean filed formalism (RMF) with various parameter
sets are compared
with experimental (where ever available) and Finite Range Droplet
Model (FRDM) prediction.
}
\renewcommand{\tabcolsep}{0.30cm}
\renewcommand{\arraystretch}{1.0}
\begin{tabular}{|c|c|c|c|c|c|c|c|c|c|c|c|c|c|}
\hline
\hline
Nucleus
& \multicolumn{8}{c|}{RMF}
&\multicolumn{2}{c|}{FRDM}
&\multicolumn{2}{c|}{Expt.}
\\
\hline
& \multicolumn{2}{c|}{NL1}
& \multicolumn{2}{c|}{NL3}
&\multicolumn{2}{c|}{NL3$^*$}
&\multicolumn{2}{c|}{FSUG}
&\multicolumn{2}{c|}{}
&\multicolumn{2}{c|}{}
\\
\hline
& PDL& NDL&PDL& NDL & PDL & NDL & PDL &NDL &PDL &NDL &PDL&NDL \\
\hline
O& 12& 29& 13&  30&12 &  30& 12&  27& 12&  26& 12& 28\# \\
Ca&  34& 69& 33&  71& 34& 71& 34& 66& 30& 73& 35\#& 58 \#\\
Ni& 49& 94& 50&  98& 50& 98& 51& 94& 46& 99& 48& 79 \\
Sn& 99& 165& 100&  172& 100& 172& 99&1 64&94&169& 99\# & 138\# \\
Pb& 178&275& 180&  281& 180& 280& 179&269&175&273& 178 & 220\# \\
114& 267& 375& 271& 392& 274& 390&  271& 376& 269& 339& 285\#&289\# \\
120& 285&376& 288&  414& 288& 410& 289& 396& 287& 339&-&- \\
\hline
\hline
\end{tabular}
\label{table3}
\end{table*}
\subsection{Proton and neutron drip-lines}

In Table \ref{table3} we have shown the proton and neutron drip-lines (PDL and NDL)
for various parameter
sets. The neutron (or proton) drip-line of an isotope is defined when the neutron (or proton)
separation energy
$S_n$ (or $S_p) \leq 0$, where $S_n=BE(N,Z)-BE(N-1,Z)$ or $S_p=BE(N,Z)-BE(N,Z-1)$ with
BE(N,Z) is the binding energy of a nucleus
with N neutron and Z proton. From the table, it is seen that all the interactions
predict almost similar proton and neutron drip-lines. If one
compares the drip-lines of NL3 and NL3$^*$, then their predictions are almost identical,
explicitly for lighter mass nuclei.  Thus, the location of drip-line with various forces
does not depend on its nuclear matter compressibility or asymmetry coefficient. For example,
the asymmetry coefficient $J=43.6$ MeV and $K_{\infty}=211.7$ MeV for NL1 set and these
are 38.68 and 271.76 MeV with NL3 parametrization. The corresponding proton drip-lines for
O isotopes are 12 and 13, and the neutron drip-lines are 29 and 30, respectively.  The similar
effects are noticed for other isotopes of the considered nuclei (see Table \ref{table3}).
\begin{figure}
\vspace{0.4cm}
\hspace{-0.3cm}
\caption{\label{fig1:gmr}The isoscalar giant monopole resonance (ISGMR)
for O, Ca, Ni and Sn isotopes from proton to neutron drip-lines
as a function of mass number. }
\includegraphics[scale=0.62]{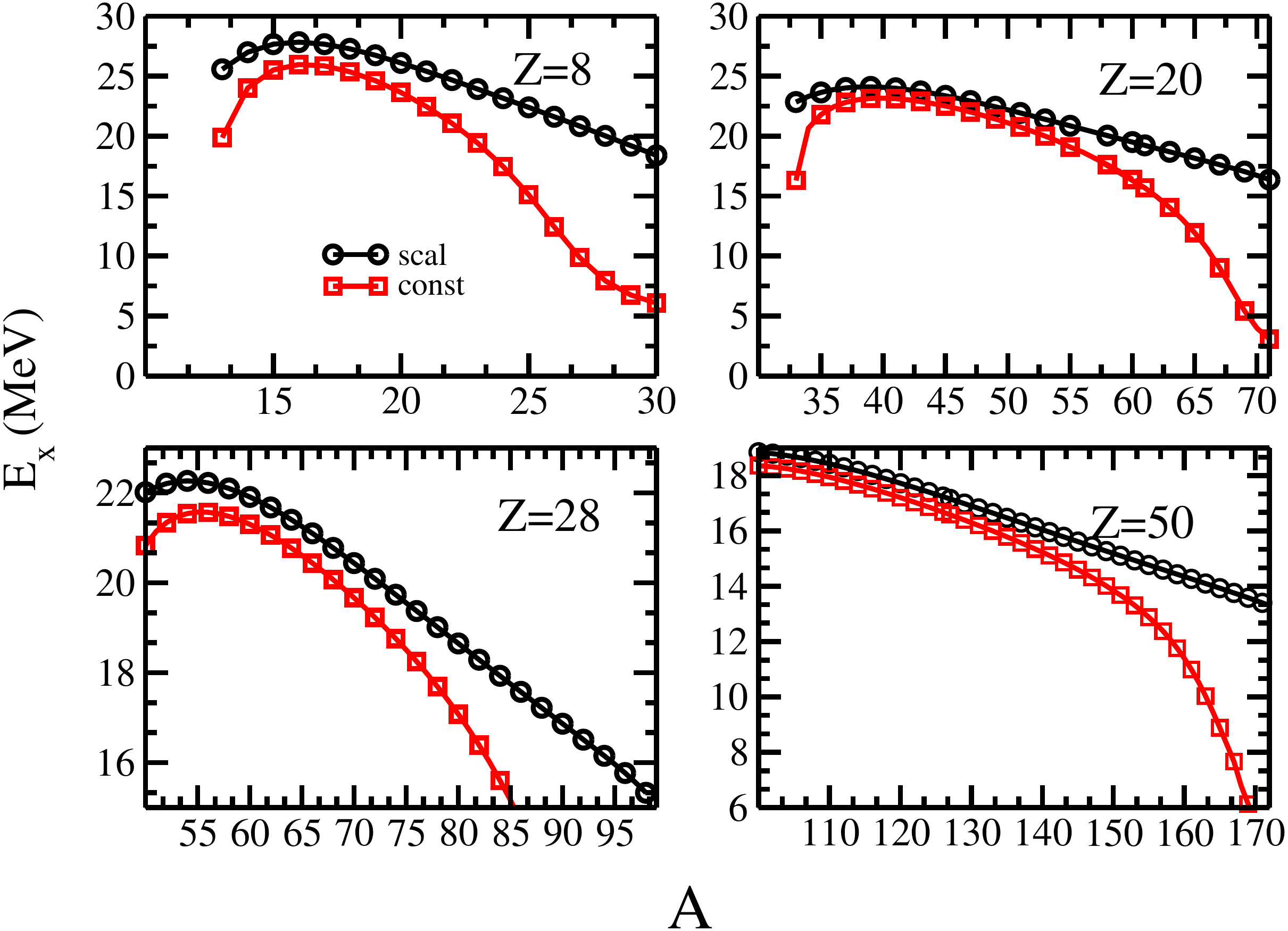}
\end{figure}

\begin{figure}
\vspace{0.4cm}
\hspace{-0.3cm}
\caption{\label{fig2:gmr}The isoscalar giant monopole resonance (ISGMR)
for Pb, Z=114 and Z=120 isotopes starting from proton to neutron drip-lines
as a function of mass number. }
\includegraphics[scale=0.62]{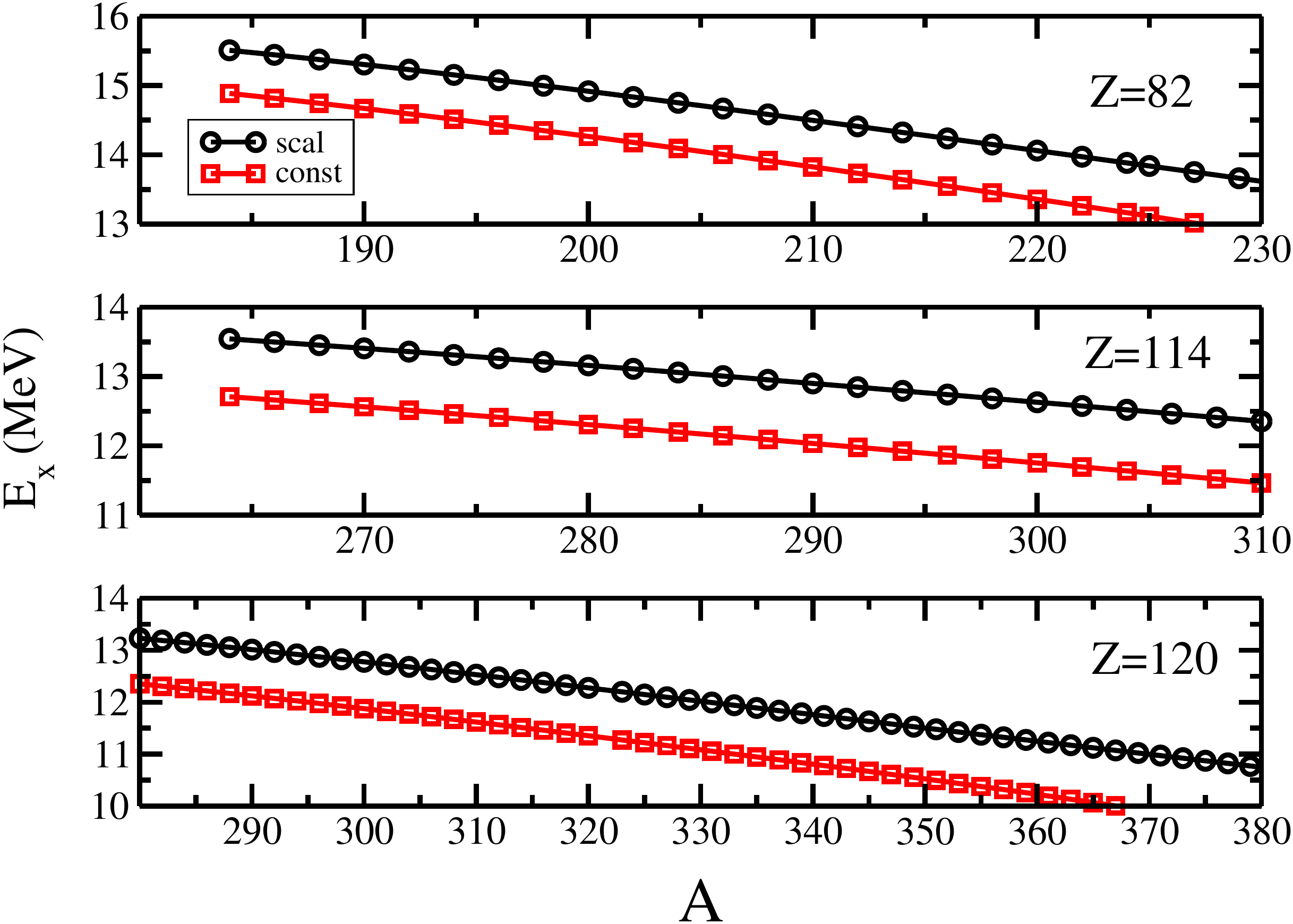}
\end{figure}
\subsection{Isoscalar giant monopole resonance}

It is well understood that the isoscalar giant monopole resonance
 has a direct relation with the compressibility of nuclear
matter which decides the softness or stiffness of an equation of
state \cite{blaizot}. This EOS also estimates the structure of neutron stars, like
mass and radius.
Thus, the ISGMR is an intrinsic property of finite nuclei as well
as nuclear equation of states and needed to be determined to gain
some light for nuclear properties. The ISGMR for O, Ca, Ni,
Sn, Pb, Z=114 and Z=120 isotopic series are given in Figs. \ref{fig1:gmr} and \ref{fig2:gmr}.
The results are calculated by using both constrained and scaling
approaches in the isotopic chain, starting from proton to neutron drip-lines.
We use the notation $E_m^s=\sqrt{\frac{AK_A^s}{B_m}}$ with the mass
parameter $B_m=\int d{\bf r}r^2{\cal H}$.
The figure shows that excitation energy  obtained from scaling calculation is always
greater than the constrained value. The difference
between the monopole excitation of scaling and constrained calculations,
generally gives the resonance width $\Sigma = \frac{1}{2}\sqrt{E_3^2-E_1^2}$,
with $E_3=\sqrt{\frac{m_3}{m_1}}$ and $E_1=\sqrt{\frac{m_1}{m_1}}$ in terms
of the ratios of the integral moments
$m_k=\int_0^{\infty}d\omega\omega^KS(\omega)$ of the RAP strength function
$S(\omega)$ \cite{mario92}. It is also equivalent to
$m_1=\frac{2}{m}A<r^2>$ and from dielectric theorem, we have
$m_{-1}=-\frac{1}{2}A(\frac{\partial R_{\eta}^2}{\partial\eta})_{\eta=0}$.

Now consider Fig. \ref{fig1:gmr}, where the excitation energy of giant isoscalar
monopole resonance $E_x$ for lighter mass nuclei are plotted. For
Z=8 the excitation energy decreases towards both proton (A=12, ${E_x}^s$= 22.51 MeV) and
neutron drip-lines (A=26, ${E_x}^s$ =21.22 MeV). This excitation energy has maximum value near
N=Z (here it is a double closed isotope with Z=8, N=8, ${E_x}^s$= 27.83 MeV). Similar trends
are followed in isotopic chain of Ca with Z=20. We find maximum excitation
energy at $^{40}$Ca (${E_x}^s={24.07}$ MeV), whereas ${E_x}^s$ is found to be smaller both in
proton (A=34, ${E_x}=23.31$ MeV) and neutron drip-lines (A=71, ${E_x}^s=16.80$ MeV).
However, the trends are somewhat different for isotopic chain of higher Z like
Z=50, 82, 114 and 120. In these series of nuclei, the excitation energy monotonically
decreases starting from proton drip-line to neutron drip-line. For example,
$^{180}$Pb and $^{280}$Pb are the proton and neutron drip nuclei having
excitation energy ${E_x}=15.63$ and  ${E_x}=11.45$ MeV, respectively.
Fig. \ref{fig2:gmr} shows clearly the monotonous decrease of
excitation energy for super heavy nuclei. This discrepancy between super heavy and light
nuclei may be due to Coulomb interaction and large value of isospin difference.
For lighter value of Z, the proton drip-line occurs at a combination of proton and
neutron where the neutron number is less than or nearer to the proton number. But for
higher Z nuclei, the proton drip-line exhibits at a larger isospin.
As the excitation energy of a nucleus is a collective property, it varies smoothly
with its mass number, which also reflects in the figures. Consider the isotopic chain
of Z=50, the drip-line nucleus (A=100) has  excitation energy 18.84 MeV and the
neutron drip nucleus A=171 has $E_x=13.39$ MeV. The difference in
excitation energy of these two isotopes is 5.32 MeV. This difference in
proton and neutron drip nuclei is 4.31 MeV for Z=82 and this is
2.37 MeV in Z=114. In summary, for higher Z nuclei, the variation of excitation
energy in an isotopic chain is less than the lighter Z nucleus. Again, by
comparing with the empirical formula of $E_x=CA^{-1/3}$, our predictions
show similar variation through out the isotopic chains. Empirically, the
value of $C$ is found to be 80 \cite{bert76}, however if we select $C=70-80$
for lighter mass isotopes and $C=80-86$ for super heavy region, then it fits well with our
results, which is slightly different than C=80 obtained by fitting the data for
stable nuclei \cite{bert76}.
\begin{figure}
\vspace{0.4cm}
\hspace{-0.3cm}
\caption{\label{fig3:sigma} The difference between the monopole excitation
energies of scaling and constrained calculations $\Sigma = \frac{1}{2}\sqrt{E_3^2-E_1^2}$
as a function of mass number A for O, Ca, Ni and Sn. }
\includegraphics[scale=0.32]{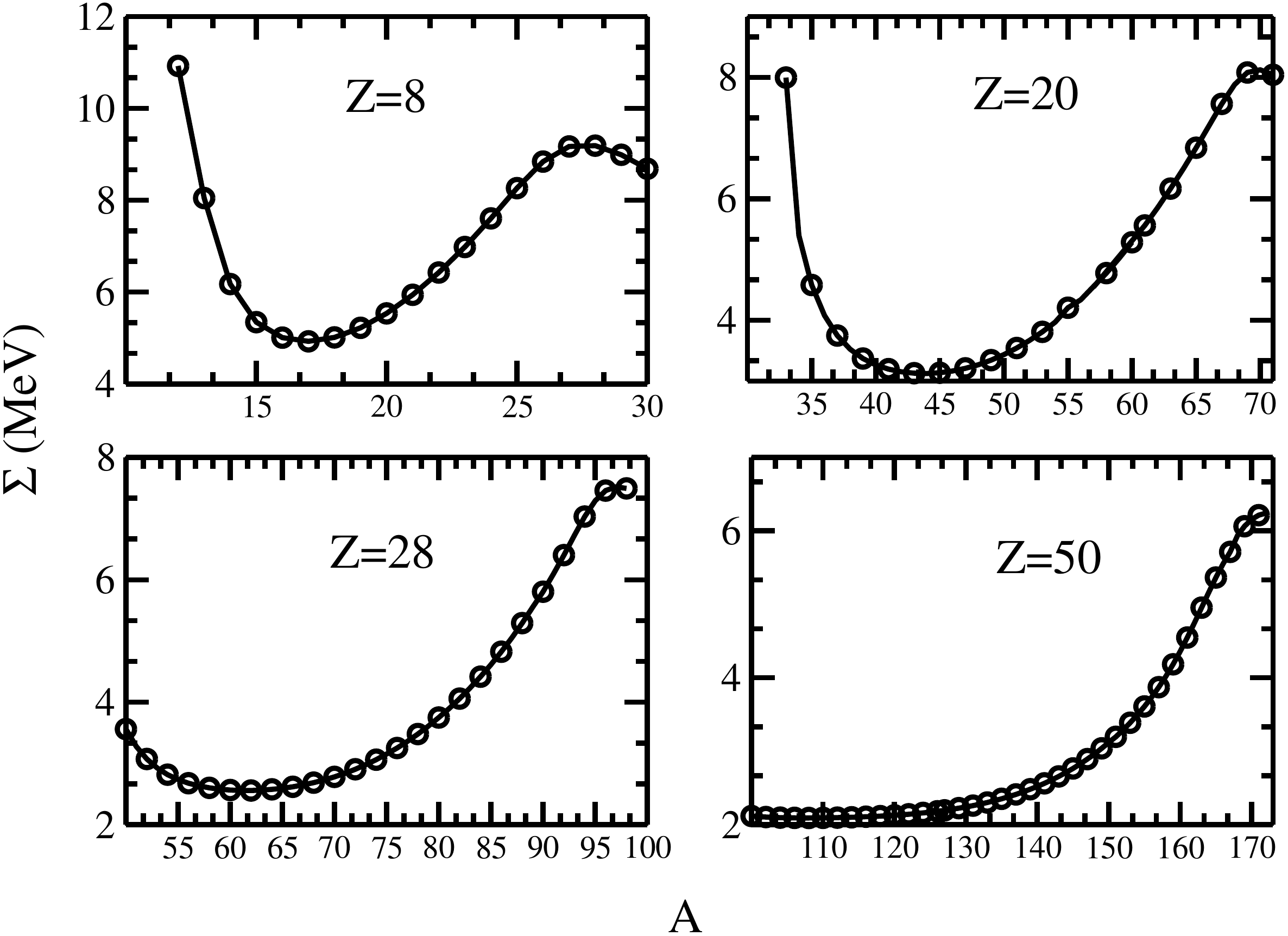}
\end{figure}

\begin{figure}
\vspace{0.4cm}
\hspace{-0.3cm}
\caption{\label{fig4:sigma} Same as Fig. \ref{fig3:sigma}, but for Pb, Z=114 and 120.
}
\includegraphics[scale=0.62]{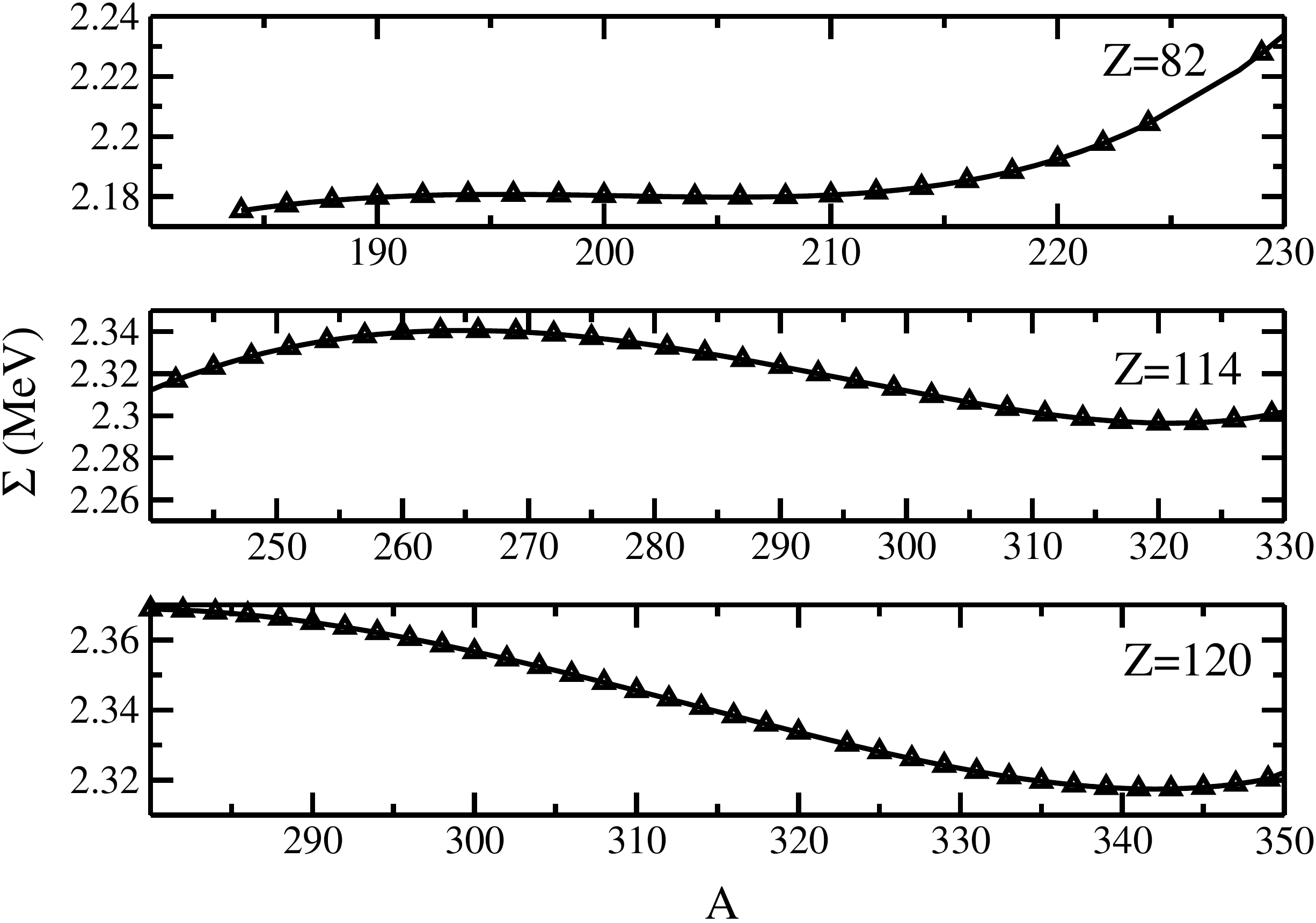}
\end{figure}
There is no direct way to calculate $\Sigma$ in the scaling or constrained
method as  random phase approximation (RPA). If we compare the excitation energy
obtained from scaling calculation with the non-relativistic RPA result,
then it is evident that the scaling gives the upper limit of the energy
response function. On the other hand, the constrained calculation predicts
the lower limit \cite{bohigas79}. As a result, the response width $\Sigma$ is obtained from
the root square difference of $E_x^s$ and $E_x^c$.
We have plotted the $\Sigma $ for the light nuclei
in Fig. \ref{fig3:sigma} and for super heavy in Fig. \ref{fig4:sigma}.
For lighter nuclei, $\Sigma$ is large both in proton and neutron drip-lines.
As one proceed from proton to neutron drip-line, the value of $\Sigma$
decreases up to a zero isospin combination (N=Z or double close) and then
increases. For example, $\Sigma$= 10.92, 5.0 and 21.62 MeV for $^{12}{O}$,
$^{16}O$ and $^{26}O$, respectively. Similar trends are also followed by
Z=20, 28 and 50 isotopic chains. This conclusion can be drawn from the results of the
excitation energy also (see Figs. \ref{fig1:gmr} and \ref{fig2:gmr}), i.e., the difference between
the scaling and constrained excitation energies are more in proton and neutron
drip-lines as compared to the Z=N region.
The value of $\Sigma$ in an
isotopic chain depends very much on the proton number. It is clear from the
isotopic chains of $\Sigma$ for O, Ca, Ni, Sn, Pb and Z=114, 120. All the
considered series have their own behavior and show various trends. Generally,
for lighter elements, it decreases
initially to some extent and again increases monotonously.
On the other hand for heavier nuclei like Pb, Z=114 and 120 this character of
$\Sigma$ with mass number is somewhat different and can be seen in Fig. \ref{fig4:sigma}.
\subsection{Compressibility modulus for finite nuclei}
\begin{figure}
\vspace{0.4cm}
\hspace{-0.3cm}
\caption{\label{fig5:comp} The compressibility modulus obtained by both
scaling and constrained approaches in the isotopic series of O, Ca, Ni and Sn.
}
\includegraphics[scale=0.62]{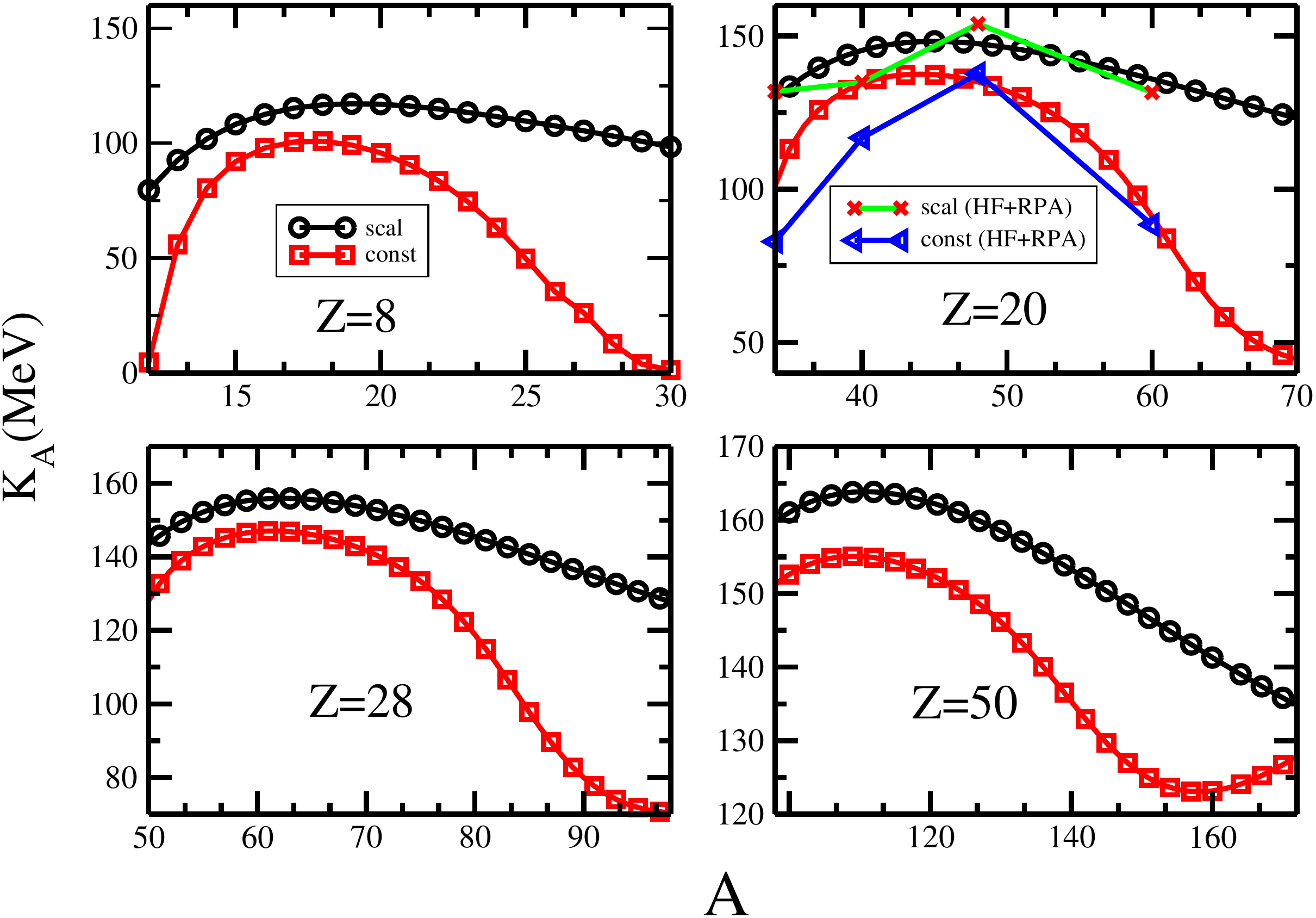}
\end{figure}

\begin{figure}
\vspace{0.4cm}
\hspace{-0.3cm}
\caption{\label{fig6:comp} Same as Fig. \ref{fig5:comp}, but for Pb, Z=114 and 120.
}
\includegraphics[scale=0.62]{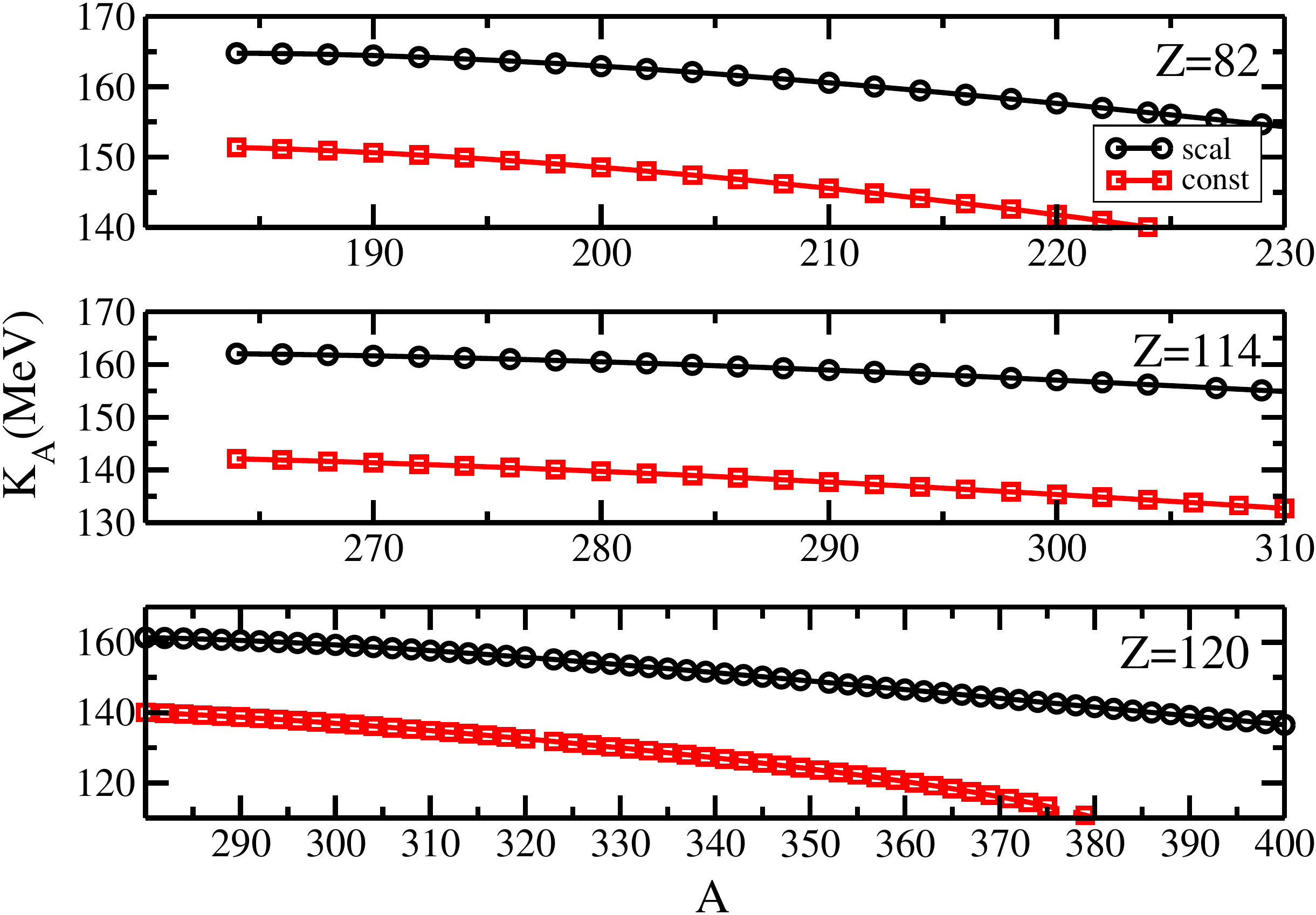}
\end{figure}
The nuclear matter compressibility $K_{\infty}$ is a key quantity in the study of equation
of state. It is the second derivative of the energy functional with respect to density
at the saturation and is defined as
$K_{\infty}=9\rho\frac{\partial^2\cal E}{\partial\rho^2}|_{\rho=\rho_0}$, which
has a fixed value for a particular force parametrization. It is well understood
that a larger $K_{\infty}$ of a parameter set, gives stiff EOS and produce
a massive neutron star \cite{aru04}. It has also
a direct relation with the asymmetry energy coefficient $J$ of the parameter set
\cite{estal01}. In the limit
A approaches to infinitely large, the finite nucleus can be approximated to
infinite nuclear matter system (N=Z for symmetry and $N\neq Z$ for asymmetry matter).
Thus, it is instructive to study the
nature of compressibility of finite nucleus $K_A$ in the isotopic chain of finite
nucleus. Here, we calculate the $K_A$ as a function of mass number for the
light nuclei considered in the present study (O, Ca, Ni, Sn) and then
extend the calculations to Pb, Z=114 and 120 in the super heavy region.
Our calculated results are shown in Figs. \ref{fig5:comp} and \ref{fig6:comp}.
The compressibility of finite
nuclei follows same trend as the excitation energy. For light nuclei, the compressibility
has small value for proton and neutron drip-lines, whereas it is maximum
in the neighborhood of double close combination.

It can be easily understood from Fig. \ref{fig5:comp} that, at a particular proton to
neutron combination,
the $K_A$ is high, i.e., at this combination of N and Z, the nucleus is more compressible.
In other word, larger the compressibility of a nucleus, it will be more compressible. Here,
it is worthy to mention that the nuclear system becomes less compressibility near both
the neutron and proton drip-lines. This is because of the the instability originating from the
repulsive part of the nuclear force, revealing a rich neutron-proton ratio, which progressively
increases with the neutron/proton number in the isotope without much affecting to
the density \cite{satpathy04}.
Similar to the excitation energy, it is found that $ K_A$ obtained by scaling method
is always higher than the constrained calculation. The decrease in
compressibility in the drip-line regions are prominent in constrained calculation than
the scaling results. From leptodermous expansion \cite{blaizot}, we can get some basic
ideas about
this decreases in the vicinity of drip-lines. The expression for finite nucleus
compressibility can be written as
\begin{eqnarray}
K_A=K_{\infty}+K_{Sur}A^{-{\frac{1}{3}}}+K_{\tau}{I^2}+K_{Coul}{Z}^2{A^{-\frac{4}{3}}},
\end{eqnarray}
where $ I={\frac{N-Z}{A}}$. The coefficient $K_{\tau}$ is negative, so compressibility
decreases with ${N-Z}$. For Ca chain, the compressibility obtained
by scaling and constrained calculations are compared with the Hartree-Fock plus
RPA method \cite{blaizot} in Fig. \ref{fig5:comp}. From Fig. \ref{fig5:comp}, one can see
that $K_A$ evaluated
by semi-classical
approximation deviates from RPA results for lighter isotopes contrary to the
excellent matching with the heavier mass of Ca isotopes.
This is because of the exclusion of the quantal correction in the semi-classical
formalism. At higher mass nuclei, this correction becomes negligible and compares to
the RPA predictions.
This result is depicted in Fig. \ref{fig6:comp} for Pb and super heavy chain of nuclei.
Here the results show completely different trends than the lighter series.
The compressibility has higher value in the vicinity of proton drip-line and
decreases monotonically towards the neutron drip-line. This is because,
for high Z-series, the proton drip-line appears at greater value of N in contrast
to the lighter mass region. Again, the compressibility decreases with
neutron number from proton to neutron drip-lines.
\begin{figure}
\vspace{0.4cm}
\hspace{-0.3cm}
\caption{\label{fig7:comp} Compressibility for finite nuclei obtained by scaling
calculation $K_A^{s}$ versus nuclear matter compressibility $K_{\infty}$.
}
\includegraphics[scale=0.62]{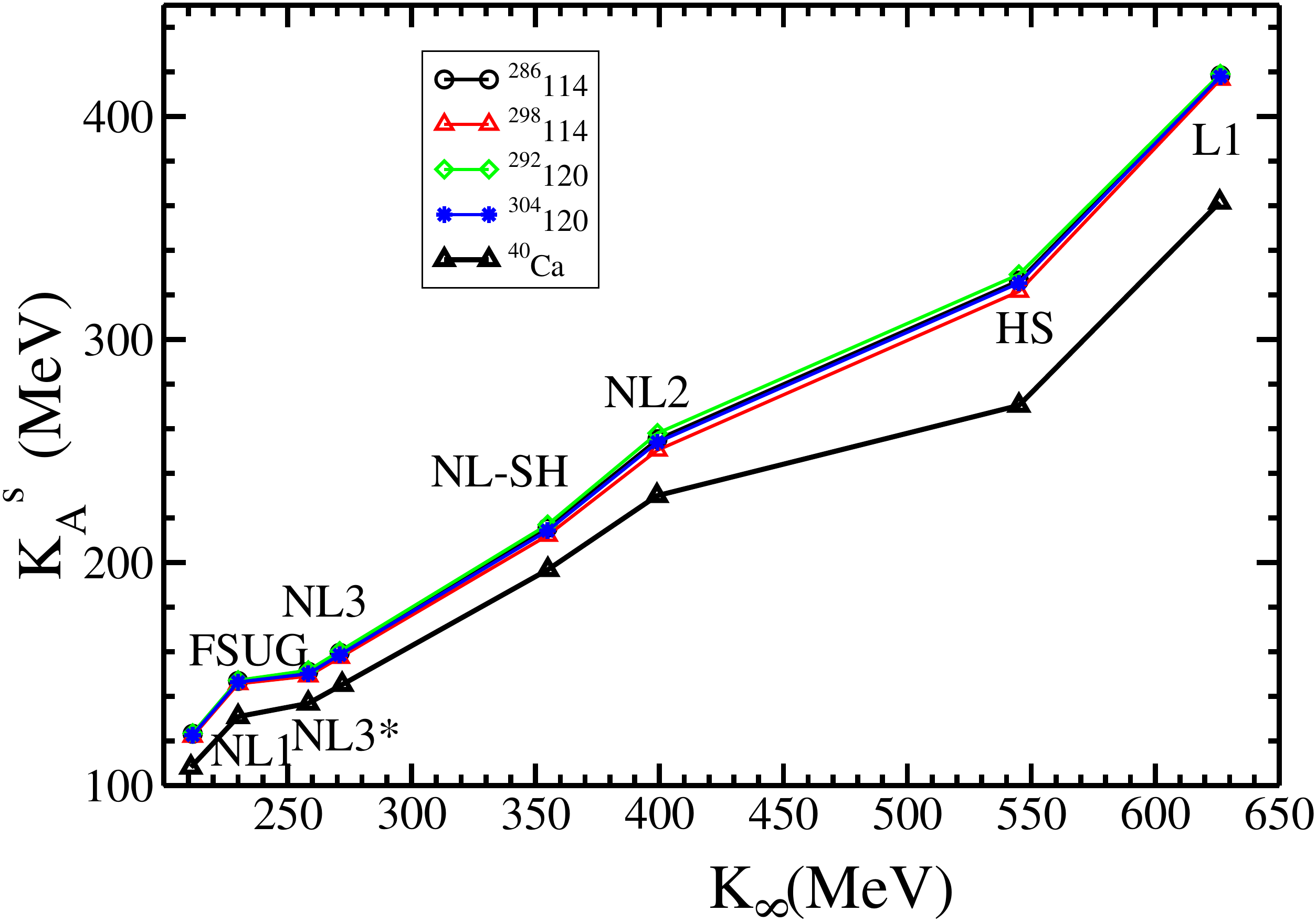}
\end{figure}

Finally, we would like to see the trend of $K_A$ with nuclear matter
compressibility for various
force parameters and also with the size of a nucleus which can reach the
infinite nuclear matter limit. For this we choose  $^{286,298}114$,
$^{292,304}120$ and $^{40}$Ca as the selected candidates and shown in
Fig. \ref{fig7:comp}. Although, the super heavy
nuclei approach the nuclear matter limit, we can not reproduce the
$K_{\infty}$ from $K_A$. This may be due to the asymmetry needed to
form a bound nucleus, which is the reason for the deviation. That means, the
asymmetry $\alpha$ of $K_A$ and $K_{\infty}$  differs significantly
(where $\alpha=\frac{N-Z}{N+Z}$),
which is the main source
of deviation of $K_A$ from  $K_{\infty}$. Also, this deviation arises due to the
surface contribution of the finite nuclei.  For a quantitative estimation,
we have calculated the $K_A^{s}$ for different force parameters having various
$K_{\infty}$ at saturation. We find almost a linear variation of $K_A^{s}$
with $K_{\infty}$ for the
considered nuclei as shown in Fig. \ref{fig7:comp}. For Ca isotopes also we find a similar nature, but smaller $K_A$ than the
super heavy nuclei.

\section{Summary and Conclusions}

In summary, we have calculated the isoscalar giant monopole resonance
for O, Ca, Ni, Sn, Pb,
Z=114 and Z=120 isotopic series starting from the proton to
neutron drip-lines. The
recently developed scaling approach in a relativistic mean field theory is
used. A simple, but accurate constrained approximation
is also performed to evaluate the isoscalar giant monopole excitation energy.
From the scaling and constrained ISGMR excitation energies, we have evaluated
the resonance width $\Sigma$ for the whole isotopic series. This is obtained
by taking the root square difference of $E_x^s$ and $E_x^c$.
The value of $E_x^s$ is always higher than the constrained result $E_x^c$.
In sum rule approach, the $E_x^s$ can be compared with the higher
and $E_x^c$ as the lower limit of the resonance width.  In general,
we found an increasing trend of $\Sigma$ for both lighter and
super heavy region near the proton and neutron drip-lines. The magnitude
of $\Sigma$ is predicted to be minimum in the
vicinity of N=Z or in the neighborhood of double close nucleus and it is maximum
for highly asymmetry systems. In the
present paper, we have also estimated the compressibility of finite nucleus.
For some specific cases, the compressibility modulus is
compared with the nuclear matter compressibility and found a linear
variation among them. It is also concluded that the
nucleus becomes less compressible with the increase of neutron or
proton number in
an isotopic chain.  Thus the neutron-rich matter, like neutron star as well as
drip-line nuclei are less compressible than the normal nuclei. In case of finite
drip-line nuclei, the nucleus is incompressible, although it possess a normal
density.

\section{Acknowledgment:} We thank Profs. X. Vi\~nas and M. Centelles for
a careful reading of the manuscript. We also thank Mr. S. K. Singh and Mr. M. Bhuyan
for discussions.


\begin{thebibliography}{99}

\bibitem{tan85}  I. Tanihata, H. Hamagaki, O. Hashimoto, Y. Shida, N. Yoshikawa,
                 K. Sugimoto, O. Yamakawa, T. Kobayashi, and N. Takahashi,
                 Phys. Rev.  Lett. \textbf{55}, 2676 (1985);
                 A. Ozawa, T. Kobayashi, T. Suzuki, K. Yoshida, and I. Tanihata
                 Phys. Rev. Lett. {\bf 84}, 5493 (2000); I. Tanihata,
                 J. Phys. G: Nucl. Part. Phys. {\bf 22}, 157 (1996).
\bibitem{han89}  P. G. Hansen and B. Jonson, Europhys. Lett.\textbf{\ 4}, 409
(1989).
\bibitem{og} Yu. Ts. Oganessian et al., Phys. Rev. Lett. {\bf 104}, 142502 (2010);
             Yu. Ts. Oganessian et al.,  Phys. Rev. Lett. {\bf 83}, 3154 (1999).
\bibitem{kumar89} K. Kumar, {\it Superheavy Elements}, (Adam Hilger, Bristol, 1989).
\bibitem{aru04} P. Arumugam, B.K. Sharma, P.K. Sahu, S.K. Patra,
     Tapas Sil, M. Centelles and X. Vi\~nas,
     Phys.  Lett. B {\bf 601}, 51 (2004).
\bibitem{young99} D. H. Youngblood, H.L. Clark and Y.-W. Lui,
                  Phys.\ Rev.\ Lett.\ {\bf 82}, 691 (1999).
\bibitem{bohigas79} O. Bohigas, A. Lane and J. Martorell,
                    Phys.\ Rep.\ {\bf 51}, 267 (1979).
\bibitem{sil04}T. Sil, S. K. Patra, B. K. Sharma, M. Centelles and X. Vi\~nas,
               Phys. Rev. C {\bf 69}, 054313 (2004); M. Bhuyan and S. K. Patra,
               Mod. Phys. Lett. A {\bf 27}, 1250173 (2012).
\bibitem{rutz95} K. Rutz, J.A. Maruhn, P.-G. Reinhard and W. Greiner,
                 Nucl. Phys. A {\bf 590}, 680 (1995).
\bibitem{sobi} P. Jachimowicz, M. Kowal and J. Skalski,
               Phys. Rev. C {\bf 83}, 054302 (2011); A. Sobiczewski, F. A. Gareev and
               B. N. Kalinkin, Phys. Lett. {\bf 22}, 500 (1966); A. Sobiczewski,
               Z. Patyk and S. C. Cwiok, Phys. Lett. {\bf 224}, 1 (1989);
               A. Sobiczewski, Acta Phy. Pol. B {\bf 41}, 157 (2010); Z. Ren,
               Phys. Rev. C {\bf 65}, 051304(R) (2002).
\bibitem{patra} M. Bhuyan, S. K. Patra and Raj K. Gupta, Phys. Rev. C {\bf 84}, 014317 (2011);
                S. K. Patra, M. Bhuyan, M. S. Mehta and Raj K. Gupta,
                Phys. Rev. C {\bf 80}, 034312 (2009); S.K. Patra,
                Cheng-Li Wu, C.R. Praharaj, Raj K. Gupta,
                Nucl. Phys. A {\bf 651}, 117 (1999).
\bibitem{patra01} S. K. Patra, X. Vi\~nas, M. Centelles and
                  M. Del Estal, Nucl. Phys. A {\bf 703}, 240 (2002);
                  {\it ibid} Phys. Lett. B {\bf 523}, 67 (2001);
                   Chaoyuan Zhu and Xi-Jun Qiu, J. Phys.\ G {\bf 17}, L11 (1991).
\bibitem{patra02} S. K. Patra, M. Centelles,
      X. Vi\~nas, and M. Del Estal, Phys. Rev. C {\bf 65}, 044304 (2002).
\bibitem{rufa86} P. G. Reinhard et al. Z. Phys. A {\bf 323}, 13 (1986).
\bibitem{boguta87}W. Pannert, P. Ring and J. Boguta, Phys. Rev. Lett.
                 {\bf 59}, 2420 (1987).
\bibitem{lala97} G. A. Lalazissis, J. K\"onig and P. Ring, Phys. Rev. C {\bf 55}, 540 (1997).
\bibitem{patra91} S. K. Patra and C. R. Praharaj, Phys. Rev. C {\bf 44}, 2552 (1991);
\bibitem{gamb90}A. R. Bodmer, Nucl.\ Phys.\ A {\bf 526}, 703 (1991);
A. R. Bodmer and C. E. Price, Nucl.\ Phys.\ A {\bf 505}, 123 (1989);
Y. K. Gambhir, P. Ring, and A. Thimet,
               Ann.\ Phys.\ (N.Y.) {\bf 198}, 132 (1990).
\bibitem{sumi93}K. Sumiyashi, D. Hirata, H. Toki and H. Sagawa,
                Nucl. Phys. A {\bf 552} 437 (1993).
\bibitem{toki94}   Y. Sugahara and H. Toki, Nucl.\ Phys.\ A {\bf 579}, 557 (1994);
S. Gmuca, J. Phys.\  G {\bf 17}, 1115 (1991); Z. Phys.\  A {\bf 342}, 387 (1992);
Nucl.\ Phys.\ A {\bf 547}, 447 (1992).
\bibitem{batty} C. J. Batty, E. Friedman, H. J. Gils and H. Rebel,
                Adv. Nucl. Phys. {\bf 19}, 1 (1989).
\bibitem{brow00}B. A. Brown,  Phys. Rev. Lett. {\bf 85}, 5296 (2000).
\bibitem{todd05} B. G. Todd-Rutel and J. Piekarewicz, Phys. Rev. Lett. {\bf 95}, 122501 (2005).
\bibitem{fattoyev10} F. J. Fattoyev, C. J. Horowitz, J. Piekarewicz and G. Shen,
                     Phys. Rev. C{\bf 82}, 055803 (2010).
\bibitem{roca11} X. Roca-Maza, M. Centelles, X. Vi\~nas and M. Warda,
                 Phys. Rev. Lett. {\bf 106}, 252501 (2011).
\bibitem{lala09} G. A. Lalazissis, S. Karatzikos, R. Fossion, D. Pena Arteaga, A. V. Afanasjev,
                 P. Ring, Phys. Lett. B {\bf 671}, 36 (2009).
\bibitem{serot86} B. D. Serot and J.D. Walecka,
                  Adv.\ Nucl.\ Phys.\ {\bf 16}, 1 (1986).
\bibitem{mario93} M. Centelles, X. Vi\~nas, M. Barranco and P. Suhuck, Ann. Phys. (NY)
                  {\bf 221}, 165 (1993).
\bibitem{mario92} M. Centeless, X. Vi\~nas, M. Barranco, S. Marco and R. J. Lombard,
                  Nucl. Phys. A {\bf 537}, 486 (1992).
\bibitem{spei98} C. Speichers, E. Engle and R. M. Dreizler, Nucl. Phys. A {\bf 562}, 569 (1998).
\bibitem{mario98} M. Centeless, M. Del Estal and X. Vi\~nas, Nucl. Phys. A {\bf 635}, 193 (1998).
\bibitem{mario93a} M. Centelles, X. Vi\~nas,  M. Barranco, N. Ohtsuka, A. Faessler,
                   Dao T. Khoa and H. M\"uther, Phys. Rev. C {\bf 47}, 1091 (1993).
\bibitem{mario10} M. Centelles, S. K. Patra, X Roca-Maza, B. K. Sharma, P. D. Stevenson
                 and X. Vi\~nas, J. Phys. G: Nucl. Part. Phys. {\bf 37}, 075107 (2010).
\bibitem{boguta77} J. Boguta, A.R. Bodmer, Nucl. Phys. {\bf 292}, 413 (1977).
\bibitem{schiff51} L. I. Schiff, Phys. Rev. {\bf 80}, 137 (1950);
{\bf 83}, 239 (1951); {\bf 84}, 1 (1950).
\bibitem{baldo13} M. Baldo, L. M. Robledo, P. Schuck and X. Vi\~nas,
                  Phys. Rev. {\bf C87}, 064305 (2013).
\bibitem{xavier11} X. Vi\~nas, P. Schuck and M. Farine
                  J. Phys. Conf. Ser. {\bf 321}, 012024 (2011).
\bibitem{maru89} T. Maruyama and T. Suzuki, Phys. Lett. B {\bf 219}, 43 (1989).
\bibitem{boer91} H. F. Boersma, R. Malfliet and O. Scholten,  Phys. Lett. B {\bf 269}, 1 (1991).
\bibitem{stoi94} M. V. Stoitov, P. Ring, and M. M. Sharma,  Phys. Rev. C {\bf 50}, 1445 (1994).
\bibitem{stoi94a} M. V. Stoitsov, M. L. Cescato, P. Ring and M. M. Sharma, J. Phys. G: Nucl. Part.
                  Phys. {\bf 20}, LI 49 (1994).
\bibitem{mari05} M. Centelles, X. Vi\~nas, S. K. Patra,
J. N. De and Tapas Sil, Phys. Rev. C {\bf 72}, 014304 (2005).
\bibitem{wang12} M. Wang, G. Audi, A. H. Wapstra, F. G. Kondev, M. MarCormick, X. Xu and
                 B. Pfeiffer, Chinese Physics C {\bf 36}, 1603 (2012).
\bibitem{angeli13} I. Angeli and K.P. Marinova, At. Data Nucl. Data Tables,
        {\bf 99}, 69 (2013).
\bibitem{bethe71} H. A. Bethe, Ann. Rev. Nucl. Sci. {\bf 21}, 93 (1971).
\bibitem{blaizot} J. P. Blaizot, Phys. Rep. {\bf 64}, 171 (1980).
\bibitem{young13} D. H. Youngblood, Y.-W. Lui, Krishichayan, J. Button,
                  M. R. Anders, M. L. Gorelik, M. H. Urin and S. Shlomo
                   Phys. Rev. C {\bf 88}, 021301 (2013).
\bibitem{bert76}F. E. Bertrand, Ann. Rev. Nucl. Part. Sc. {\bf 26}, 456 (1976);
          J. Speth, A. V. Woude, Rep. Porg. Phys. {\bf 44}, 719 (1981);
          K. Goeke, J. Speth, Ann. Rev. Nucl. Part. Sc. {\bf 32}, 65 (1982).
\bibitem{estal01} M. Del Estal, M. Centelles, X. Vin˜as, and S. K. Patra,
                  Phys. Rev. C {\bf 63}, 024314 (2001).
\bibitem{satpathy04} L. Satpathy and S. K. Patra,  J. Phys. G:
                     Nucl. Part. Phys. {\bf 30}, 771 (2004);
                     S K Patra, R K Choudhury and L Satpathy,  J. Phys. G:
                     Nucl. Part. Phys. {\bf 37}, 085103 (2010).





\end{thebibliography}
\end{document}